\documentclass[]{emulateapj}
\usepackage{natbib}
\usepackage{apjfonts}
\usepackage{mathptmx}

\def\simlt{\vcenter{\hbox{$<$}\offinterlineskip\hbox{$\sim$}}}
\def\simgt{\vcenter{\hbox{$>$}\offinterlineskip\hbox{$\sim$}}}

\def\Bb{$B_{435}$\/}

\def\Vb{$V_{606}$\/}

\def\zb{$z_{850}$\/}
\def\ib{$i_{775}$\/}

\def\zbAB{$z_{850,\rm AB}$\/}

\def\Vi{$V_{606}-i_{775}$}
\def\iz{$i_{775}-z_{850}$}

\def\AaC{{\em A}~and~{\em C}}
\def\AC{{\em A--C}}

\def\OII{[O~II]$\lambda$3727}

\def\O5007{[O~III]$\lambda$5007}
\def\Mg2{Mg~II}

\def\sfr1600{SFR$_{1600}$}
\def\rhoz{$\rho_\mathrm{SFR}(z)$}

\slugcomment{Accepted for publication in the Astronomical Journal}

\shorttitle{\sc The Morphological Demographics of galaxies}
\shortauthors{\sc F. Menanteau et~al.}

\begin{document}

\title{The Morphological Demographics of galaxies in the ACS Hubble Ultra Deep Parallel Fields\altaffilmark{\dag}}

\author{
\centerline{
Felipe~Menanteau\altaffilmark{1}, 
Holland~C.~Ford\altaffilmark{1},
Ver\'onica~Motta\altaffilmark{2,1},
Narciso~Ben\'{\i}tez\altaffilmark{3,1}
}
Andr\'e~R.~Martel\altaffilmark{1},
John~P.~Blakeslee\altaffilmark{1}
and
Leopoldo~Infante\altaffilmark{2}
}

\altaffiltext{1}{Department of Physics and Astronomy, Johns Hopkins
University, 3400 North Charles Street, Baltimore, MD 21218.}
\altaffiltext{2}{Departmento de Astronom\'{\i}a y Astrof\'{\i}sica,
Pontificia Universidad Cat\'olica de Chile, Casilla 306, Santiago
22, Chile.}
\altaffiltext{3}{Instituto de Astrof\'{\i}sica de Andaluc\'{\i}a (CSIC),
Camino Bajo de Hu\'etor 24, Granada 18008, Spain }
\altaffiltext{\dag}{Based on observations obtained with the Hubble
Space Telescope and at Las Campanas Observatory.}

\begin{abstract}

We present a morphological analysis of distant field galaxies using
the deep ACS images from the public parallel NICMOS observations of
the Hubble Ultra Deep Field obtained in the F435W (\Bb), F606W (\Vb),
F775W (\ib) and F850LP (\zb) filters. We morphologically segregate
galaxies using a combination of visual classification and objective
machine based selection. We use the Asymmetry ($A$) and Central
Concentration ($C$) parameters to characterize galaxies up to
\zbAB$<25$~mag. We take advantage of the multicolor dataset and
estimate redshifts for our sample using the Bayesian photometric
redshift (BPZ) which enables us to investigate the evolution of their
morphological demographics with redshift. Using a template fitting model
and a maximum likelihood approach, we compute the star-formation rate
(SFR) for galaxies up to $z\simeq1.3$ and its contributions from
different morphological types. We report that spirals are the main
providers to the total SFR. The E/S0s contribution flattens out at
$z\simeq1$ while the Irr/Pec populations continuously rise to match
the spirals contribution at $z\simeq1.0$. We use the \iz\ and \Vi\
color-magnitude diagrams to constrain the galaxies' formation histories
and find that E/S0s show both a population of luminous red galaxies in
place at $z\sim1.2$ and a bluer and fainter population resembling
those of Irr/Pec at similar redshifts.

\end{abstract}


\section{Introduction}

The archeological nature of galaxy evolution studies prevents us from
following individual galaxies over time. In consequence we are only
left with snapshots at different lookback times from where we attempt,
as excavators, to piece together their evolutionary histories from the
motion of different observable quantities and their
inter-relationships through different cosmic times. As morphology
correlates with a range of physical properties in galaxies, such as
mass, luminosities and particularly color, this suggest that their
appearance must embody some important clues about their formation
histories. Moreover, the growing acceptance of the notion that the
morphological appearance of galaxies may not be a static property set
at an early stage of formation, makes it crucial to understand the
flow of the morphological mix of galaxies and their observables as a
function of redshifts.

The early morphological studies of HST galaxies
\citep{Cowie-etal-95,van-den-Bergh-etal-96,Volonteri-etal-00,Reshetnikov-etal-03}
set the stage for the Advanced Camera for Surveys (ACS;
\citealt{Ford-etal-02}) to effortlessly resolve galaxies at ever
fainter limits thus enabling the study of the morphological demographics of
distant galaxies with larger and deeper datasets \citep[see][for a
list of recent studies]{Elmegreen-Elmegreen-Hirst-04,
Elmegreen-Elmegreen-Seets-04, Elmegreen-Elmegreen-Rubin-05}.  Since
its installation, ACS has provided a continuous flow of high
resolution optical imaging of distant field galaxies over
significantly wider areas. Large programs such as GOODS
\citep{Giavalisco-etal-04a}, GEMS \citep{Rix-etal-04} and the ACS/GTO
\citep{Postman-etal-05,Ford-etal-02} have spawned several studies that
address the color evolution of galaxies, the number evolution of red
objects with redshift \citep{Bell-etal-04}, the mass assembly rate for
different morphologies \citep{Bundy-etal-05} and constrain the ages
and masses of early types \citep{Treu-etal-05,Menanteau-etal-04}.

In this paper we take advantage of public deep multicolor ACS parallel
observations to study the morphological color and star formation rate
evolution of galaxies at depths comparable to those of the Hubble Deep
Fields \citep{Williams-etal-96}, but probing areas many times larger.
Throughout this paper we use $H_0=70$~km~s$^{-1}$Mpc$^{-1}$ and a flat
($\Omega_k=0$) cosmology with $\Omega_M=0.3$ and
$\Omega_{\Lambda}=0.7$. Magnitudes are given in the AB system across
the paper.

\section{Observations}

\subsection{ACS Multicolor Imaging}

\begin{figure}
\plotone{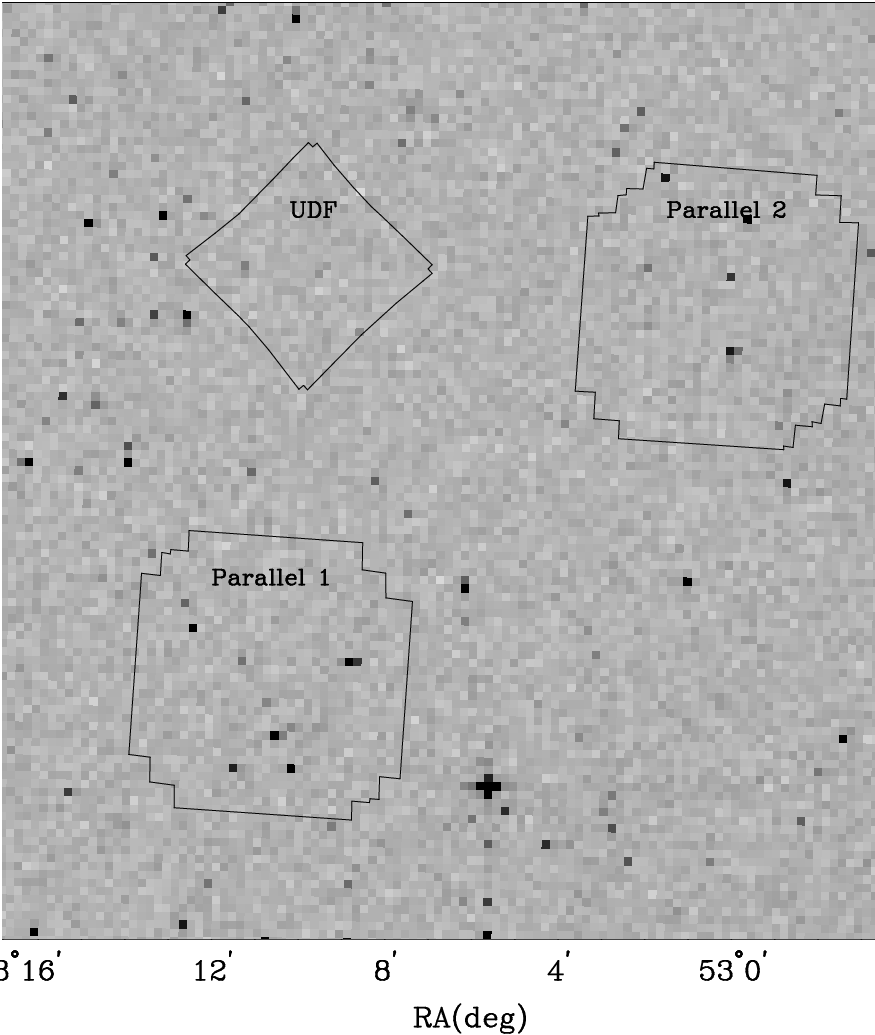}
\caption{The diagram of the relative positions and sizes of the final
mosaiced patterns of the ACS parallel observations of the UDF NICMOS
and the ACS/UDF observations.}
\label{fig:diagram}
\end{figure}

Our analysis is based on the deep ACS parallel fields of the Near
Infrared Camera and Multi-Object Spectrometer (NICMOS) imaging of the
Hubble Ultra Deep Field (UDF) (GO 9803, PI: Thompson). The ACS
datasets consist of two separate fields around the center of the UDF
arranged in a mosaiced pattern determined by the NICMOS
observations of $~30''$ spacing between pointings
\citep{Blakeslee-etal-04}. The resulting coverage of the parallel ACS
fields as well as the UDF and NICMOS observations positions are shown
in Fig~\ref{fig:diagram}. Each field comprises 9, 9, 18, and 27 orbits
in the F435W ($B_{435}$), F606W ($V_{606}$), F775W ($i_{775}$) and
F850LP ($z_{850}$) filters respectively.
The images are publicly available from the STScI archive and were
processed using the ACS/GTO pipeline (Apsis;
\citealt{Blakeslee-etal-03}) which includes object detection,
photometry and cataloging using SExtractor \citep{SEX}. Photometry was
calibrated in AB magnitudes using the zeropoints from Sirianni
et~al. (2005). Photometric redshifts were also computed using the
Bayesian estimator BPZ from \cite{Benitez-00}. Within the deepest
regions of the fields the $10\sigma$ limited magnitudes were 28.8,
29.0, 28.5 and 27.8 in \Bb, \Vb, \ib\ and \zb\ respectively
\citep{Bouwens-etal-04} in a scale of
0\farcs05~pixel$^{-1}$. 
The final ACS area covered was 55.7 arcmin$^2$ from both
fields. For a detailed description of the dataset and its processing
we refer to \cite{Blakeslee-etal-04}.

\subsection{LDDS2 Magellan Spectra}

Follow up multi-object spectroscopy on one of the parallel fields was
performed using the low-dispersion survey spectrograph 2 (LDSS2) on
Las Campanas Observatory Magellan Clay 6.5m Telescope. A complete
description of the observations and data reduction analysis of the
spectroscopic sample is given in \cite{Motta-etal-05}; here we briefly
summarize these observations. For Field \#1 three masks were observed
in November 25 and 26, 2003 from which spectra for a total of 56
galaxies were secured. From these we secured redshifts for 51, and for
a subset of 37 galaxies line strengths and equivalent widths (EW) were
measured. Targets for the masks were uniformly selected from galaxies
within the field with \ib$<23$.  The spectra were reduced using a
combination of a customized IDL reduction software techniques
optimized for the extraction of background limited data
\citep{Frye-etal-02} and IRAF tasks for the wavelength and flux
calibrations.
The final integrated exposure times per mask were between 7200s and
4800s at a central wavelength of 5500\AA\ and spectral coverage of
$4000-9000$\AA\ at 5.3\AA~pixel$^{-1}$ dispersion.

\subsection{Morphological Classification}
\label{sec:morph}

Rather than using the color of galaxies as a surrogate for
morphological type, in this paper we exploit the superb ability of ACS
to resolve and morphologically segregate distant galaxies with no a
priori built-in biases in the selection criteria.
We classified galaxies according to their morphologies down to
\zbAB$<25$ using a two-pronged approach based on visual inspections
and automated classifications. First, galaxies were visually inspected
and classified using the Medium Deep Survey (MDS) scheme as described
in \cite{Abraham-etal-96} and modified by \cite{Menanteau-etal-99}
\citep[see~also][]{Treu-etal-05}. Within this system we segregated
galaxies into three broad categories: $i)$ E/S0 which include E, E/S0
and S0; $ii)$ Spirals: including Sa+b, S and Sc+d and $iii)$ Irr/Pec
including Irregular and Peculiar types. We also limited our sample to
objects with $>90\%$ confidence in their BPZ redshift estimates. We
elaborate on this in the next section. A final catalog containing 1228
galaxies was visually inspected by one of us (FM) in both ACS fields.

Secondly, we independently investigated the morphologies of galaxies
using automated classifications. For this we calculated the central
concentration ($C$) and asymmetry ($A$) parameters for all galaxies
using the definitions from \cite{Abraham-etal-94,Abraham-etal-96}.
Our analysis is based on the well-known relationship between the $A$
and $C$ structural parameters and traditional visual morphologies
\citep{Abraham-etal-96-HDF,Abraham-etal-96,Brinchmann-etal-98,Conselice-03}
We compute asymmetry and concentration using {\tt PyCA}, our own
automated Python software. The code is designed to compute \AaC\ from
the SExtractor products generated by Apsis, the detection catalog
and associated segmentation image. However, it can be used with any
standard SExtractor products and it is publicly
available\footnote{It can be downloaded from
http://acs.pha.jhu.edu/$\sim$felipe/PyCA}.

Concentration is computed as the ratio between the flux at 30\% of a
given radius $0.3R$ to the total flux up to that radius $R$. Formally
it is defined as,
\begin{equation}
C = f(0.3R)/f(R), 
\end{equation}
were $f(R) = 2\pi \int_0^R I(r)dr$ is the integrated flux within the
radius $R$. The determination of the radius $R$ up to where to integrate
the light is of importance particularly for the asymmetry values. 
As we sample galaxies over ever larger cosmic ages, the surface
brightness of galaxies at higher redshifts become heavily affected by
the well-known $(1+z)^4$ cosmological dimming
\citep[][]{Tolman-1934,Sandage-Lubin-01}.  Therefore to consistently
measure structural parameters of galaxies over similar physical areas
at different redshifts, we chose an aperture radius defined as $R_{\rm
p}=1.5\times r_{\rm p}$, with $\eta=0.2$, where $r_{\rm p}$ is the
\cite{Petrosian-76} radius and
$\eta(r) = I(r)/\langle I(r) \rangle$
is the ratio of the galaxy average surface brightness, $I(r)$, in an
annulus of radius $r$ and the mean value up to the same radius,
$\langle I(r) \rangle$ \citep[see][and references
therein]{Papovich-etal-03, Blanton-etal-01, Menanteau-etal-04}. The
advantage of using a Petrosian radius over traditional surface
brightness limits ones, such as the Holmberg radius
\citep[i.e.][]{Abraham-etal-96-HDF,Menanteau-etal-99} is that it only
depends on the galaxy light profile and it is therefore independent of
the redshift of observation \citep[see][for a
discussion]{Bershady-etal-00}. Comparatively accurate results can be
obtained using the SExtractor Kron radius, which in most cases
coincides with $R_{\rm p}$.

Similarly, asymmetry is computed as half of the ratio between the
absolute value of subtraction of the image, $I_{ij}$, and its
$180^{\circ}$ rotated image around its center, $I_{ij}^R$, to the
original image, both contained in an elliptical mask of semi-major length
$R_{\rm p}$. To avoid variations arising from noise features, the
pivot point of the galaxies was computed after the images were
smoothed with a Gaussian kernel of 1~pixel and then
self-subtracted.
Formally, our working definition for
asymmetry over a galaxy with an intensity matrix $I_{ij}$ can be
expressed as,
\begin{equation}
A = \frac{1}{2}\frac{\sum |I_{ij} - I_{ij}^R| - b}{\sum I_{ij} }.
\end{equation}
Because we use the absolute values of the self-subtracted galaxy
residuals, small variations in the image background can appear even in
the most symmetric objects, consequently we add a factor $b$ to
correct for this artificial $A$ signal introduced, such as
$b=\sqrt{2}\sigma_{\rm sky}N_{\rm pix}$, where $N_{\rm pix}$ is the
area inside the ellipse mask and $\sigma_{\rm sky}$ is the sky mean
standard deviation.

\begin{figure}[t]
\centerline{
\includegraphics[width=3.42in, angle=0]{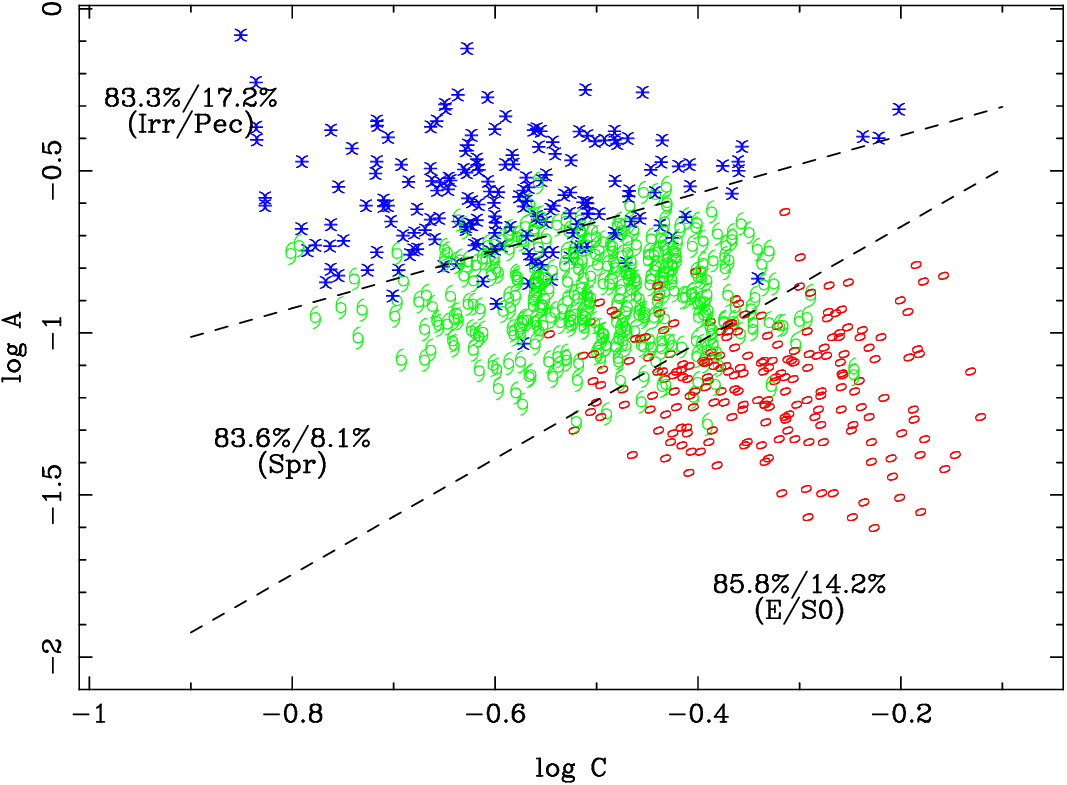}}
\caption{The asymmetries and central concentrations values keyed to
their visual morphological classes for all 768 objects in the Field
\#1. E/S0 systems are represented by ellipses; Spiral galaxies by
spirals and Irr/Pec as asterisks. The dashed lines represent the
selection limits based on $A-C$ that maximizes the recovery of visual
classes and minimizes the contamination from other morphological
types. Numbers inside the figure represent the recovery and
contamination percentages of visual classes and from other classes
respectively using the \AaC\ based selection.}
\label{fig:eye-AC}
\end{figure}

We compute $A$ and $C$ for all objects in our galaxy sample to a
limiting magnitude \zbAB$<25$. To avoid morphological variations
resulting from redshifting bandpass (i.e. morphological
$K-$correction), we compute $A$ and $C$ using the image in the closest
available band to the rest-frame $B$ derived from the photometric
redshift information. In Fig.~\ref{fig:eye-AC} we show the
distribution of asymmetry and concentration for galaxies in Field \#1
as function of their visual morphological classes. We only show
structural values for one field to avoid overcrowding in the
diagram. Results are similar for Field \#2. This type of diagram has
been extensively used to illustrate and calibrate the morphological
segregation of galaxies based purely on machine classifications
\citep[see][]{Abraham-etal-96-HDF,Abraham-etal-96,Menanteau-etal-99}.
We use this approach to isolate galaxies in the same three equivalent
broad visual morphological categories. The \AC\ classification recovers
~$\simgt80\%$ of the visual classifications in all three categories,
with contaminations of $\simlt15\%$ from other classes. We will use
this criteria to isolate galaxies based on $A-C$ and compare their
properties in the next section.

\section{Analysis}

One of the objectives in this paper is to investigate the cosmic star
formation history and its dependence on morphological type. To
accomplish this we need to obtain reliable SFRs for a large number of
galaxies. Because of the faint limits of the sample and the large
number of objects involved, securing spectroscopic diagnostic for all
of them would be extremely time consuming, and in most cases
unfeasible because of their faintness. Instead, we chose to follow an
approach that takes advantage of the deep and precise ACS multicolor
photometric information, and we use ancillary ground-based
spectroscopy and UV imaging for a control sample of galaxies to
validate our results.

\subsection{The UV SFR at 1600\AA}

\begin{figure}
\centerline{
\includegraphics[width=3.45in, angle=0]{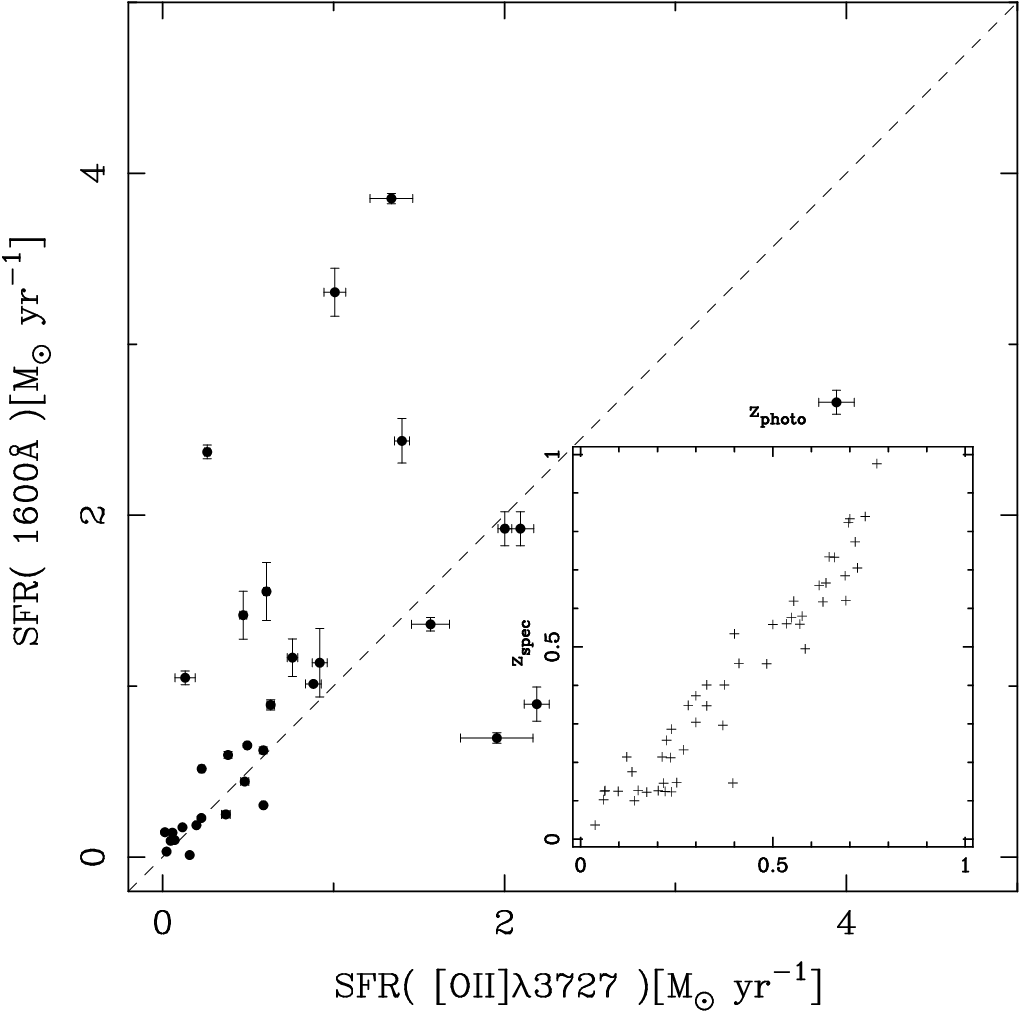}}
\caption{The star formation rate (SFR) from the spectroscopic sample
of galaxies derived from the \OII\ EW compared to the SFR estimation
based on the 1600\AA\ UV flux. The inset panel shows the relation
between the BPZ photometric redshifts estimates and the spectroscopic
ones from our ground-based observations.}
\label{fig:OII-1600}
\end{figure}

There are several SFR indicators based on transformation between
ultraviolet flux and star formation activity \citep[see][for a
discussion and references
therein]{Steidel-etal-99,Coe-etal-05,Hopkins-etal-01,Hopkins-etal-04}. We
opt for computing the SFR for each galaxy using the prescription of
\cite{Meurer-etal-99}. This is straightforward and computationally
easy to implement, and derived from a constant SFR and a Salteper
initial mass function with limits $0.1-100~M_{\sun}$. They define the
ultraviolet flux at 1600\AA, $F_{1600}$, as the integrated light over
a square synthetic filter of width 350\AA\ and central rest-frame
wavelength at 1600\AA. Based on this, \cite{Meurer-etal-99} provide a
direct transformation between a galaxy's SFR and $F_{1600}$ such as
\[
{\rm SFR}[{\rm M}_{\sun}{\rm yr}^{-1}]=10^{-0.4(M_{1600}+18.15)}
\]
and $M_{1600}$ is the $F_{1600}$ corresponding absolute magnitude in
AB system.

Because our ACS filter information only spans between
$\sim3500-9000$\AA, the galaxies' rest-frame UV light only falls into
our bluest band for $z>1.5$. However, due to the large number of
objects involved in this study and their precise photometry it is
possible to derive a general description the galaxies' spectral energy
distribution (SED). We utilize the galaxy's reconstructed SED from
their BPZ spectral type, $T_B$, and based on this we estimated the
galaxy's UV light at 1600\AA\ and compute their rest-frame $F_{1600}$
flux. The $T_B$--types provided by BPZ are a linear combination from a
template library of SEDs. We use the CWWSB\_Benitez2003 template set
as described in \cite{Benitez-etal-04}. These are based on the
templates from \cite*{CWW-80} and \cite{Kinney-etal-96} consisting of
El, Sbc, Scd, Im, SB3, and SB2, which represent the typical SEDs of
elliptical, early/intermediate type spiral, late-type spiral,
irregular, and two types of starburst galaxies respectively. This
template set has been modified from earlier BPZ versions to remove
differences between the predicted colors and those of real galaxies,
which results in improved BPZ estimates \cite[see][for
details]{Benitez-etal-04}. We note that \cite{Coe-etal-05} have
followed a similar approach to estimate the UV-SFR using also the
template fits from BPZ for $z<6$ galaxies in the HUDF, but at 1400\AA\
instead.

However, because new stars are born in dusty regions, their observed
UV light can be strongly attenuated and our observed UV SFRs are
therefore subject to obscuration. In order to account for this we must
correct the SFR estimates for dust extinction. We use the empirical
relationship between the unobscured far-infrared (FIR) flux and the
intrinsic SFR (SFR$_i$) prescription by \cite{Hopkins-etal-01}, as FIR
light penetrates the dust and provides reliable SFR$_i$.  The
\citeauthor{Hopkins-etal-01} empirical formulation between the
(obscured) observed-UV SFR, SFR$_o$, at a given wavelength $\lambda$
and their SFR$_i$ is,

\begin{displaymath}
\log(\mbox{SFR}_i) = \log(\mbox{SFR}_o)-0.346 \times k(\lambda)
\end{displaymath}
\begin{equation}
\hspace{3.0cm}\times\log \left[\frac{0.797\log (\mbox{SFR}_i) + 3.834}{2.88}\right],
\end{equation}
where SFRs are in units of [M$_\sun$yr$^{-1}$] and $k(\lambda)$ is the
empirical \cite{Calzetti-etal-00} attenuation law,
$k(\lambda)=A(\lambda)/E(B-V)$ This improper equation must be solved
numerically to obtain the extinction-corrected SFR$_i$ for a given
observed UV SFR$_o$. We compute the corrected SFR for our full sample
for $\lambda=1600$\AA\ using $k(1600)=9.97$. This same method has also
been used in the HUDF for high redshift galaxies (see Coe et~al. 2005
and references therein). In the next section we use the corrected
SFR$_i$ to calculate the star formation density evolution as a
function of morphological type.

\subsection{Emission Line vs. Synthetic SFR}

We compare our SFR estimates obtained using the 1600\AA\ flux with
independent SFR measurement from traditional EWs.  For this, we
utilize the EW from the \OII\ emission lines to estimate SFR using the
prescription from \cite{Kennicutt-98} using the control-sample of
galaxies from our LDDS2 spectroscopic campaign. In
Fig~\ref{fig:OII-1600} we show the individual \sfr1600\ estimates from
our synthetic approach and the ones obtained from \OII\ EWs. For the
majority of galaxies there is good agreement between the two
measurements, showing a clear relation between both estimators.
We find that although there are minor discrepancies for individual
galaxies, the overall agreement between \sfr1600 and \OII\ is quite
satisfactory and we estimate out typical uncertainties to be $\Delta{\rm
SFR}/{\rm SFR}\sim0.3$. As in this paper we target the general
properties for the population as a whole rather than the precise
characterization of each galaxy, we are confident that our
template-based SFR yield meaningful results.

\subsection{Comparing Synthetic UV fluxes with GALEX observations}

\begin{figure}[t]
\centerline{
\includegraphics[height=1.7in, angle=0]{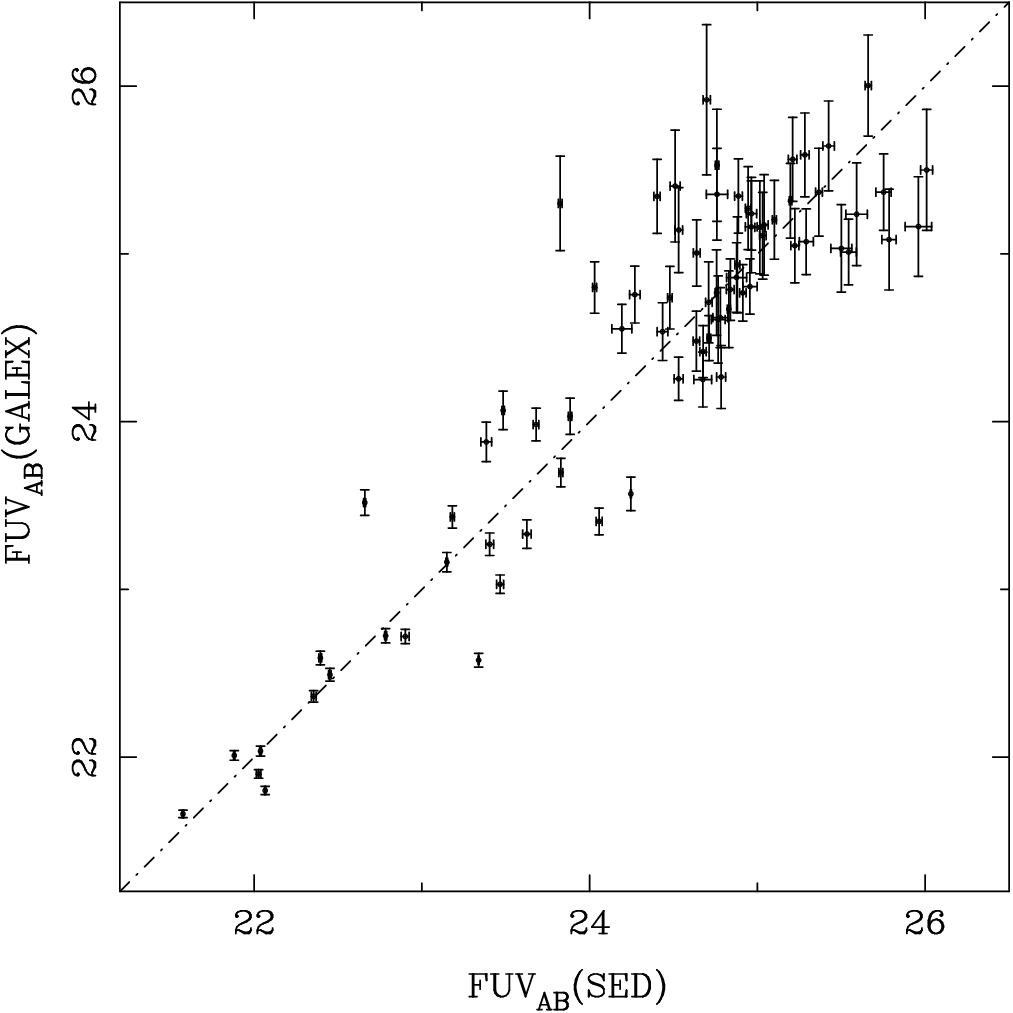}
\includegraphics[height=1.7in, angle=0]{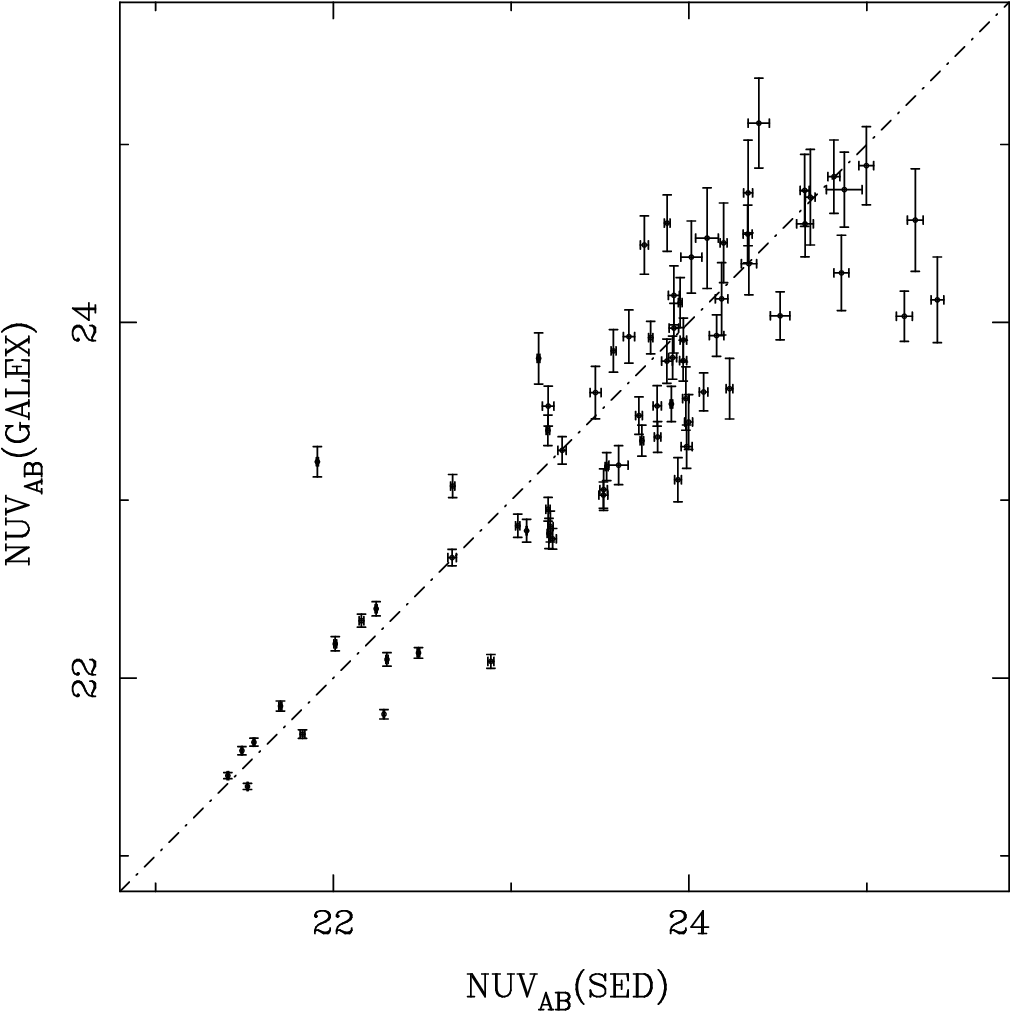}}
\caption{The GALEX observed FUV (left panel) and NUV (right panel)
  magnitudes compared to the synthetic recovered values using the
  reconstructed spectral energy distributions from the BPZ template
  library set. GALEX FUV and NUV error bars come from the original
  errors in the GALEX catalogs. Synthetic error bars reflect only
  original ACS photometric errors.}
\label{fig:galex}
\end{figure}

We further investigate our ability to model the galaxies' UV light by
comparing our synthetic estimates of the UV flux with the observed
magnitudes obtained from overlapping Galaxy Evolution Explorer (GALEX,
\citealt{Martin-etal-05}) observations for the same ACS regions. We
retrieved the positions, FUV(1530\AA) and NUV (2310\AA) magnitudes and
associated errors in the AB system from the MAST/GALEX Web
Archive\footnote{http://galex.stsci.edu/GR1} and matched them with the
galaxies in our ACS sample. From a total of 165 GALEX sources in both
fields, we successfully identified 71 galaxies in our sample with both
FUV and NUV magnitudes. We compute the synthetic magnitudes through
the FUV and NUV bandpasses for each of the matched galaxies using the
reconstructed SEDs. In Figure~\ref{fig:galex}, we show the comparison
of real and synthetic values of FUV and NUV. We find that the real and
modeled values agree well, with a nominal mean scatter $\Delta(UV_{\rm
GALEX}-UV_{\rm SED})$ of 0.31 and 0.36 mags for FUV and NUV
respectively. Therefore, we conclude that our UV flux estimates for
ACS galaxies represent a good description of their real values and can
be used as surrogates when measuring the SFR for large samples of
galaxies.

\section{Morphological demographics between $0.3<z<1.2$} 

\subsection{The morphological SFR density}

\begin{figure*}
\centerline{
\includegraphics[height=3.45in, angle=-90]{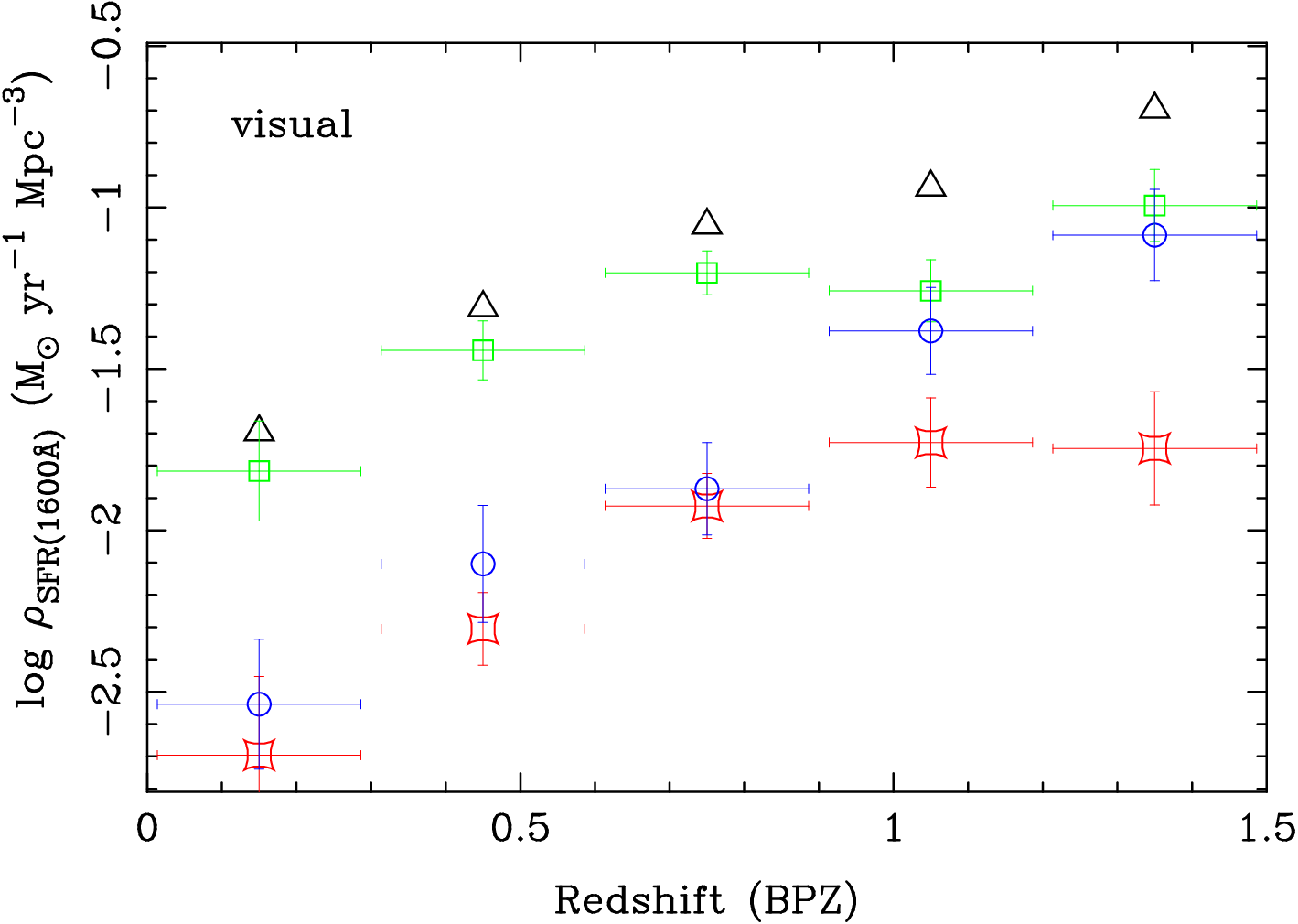}
\includegraphics[height=3.45in, angle=-90]{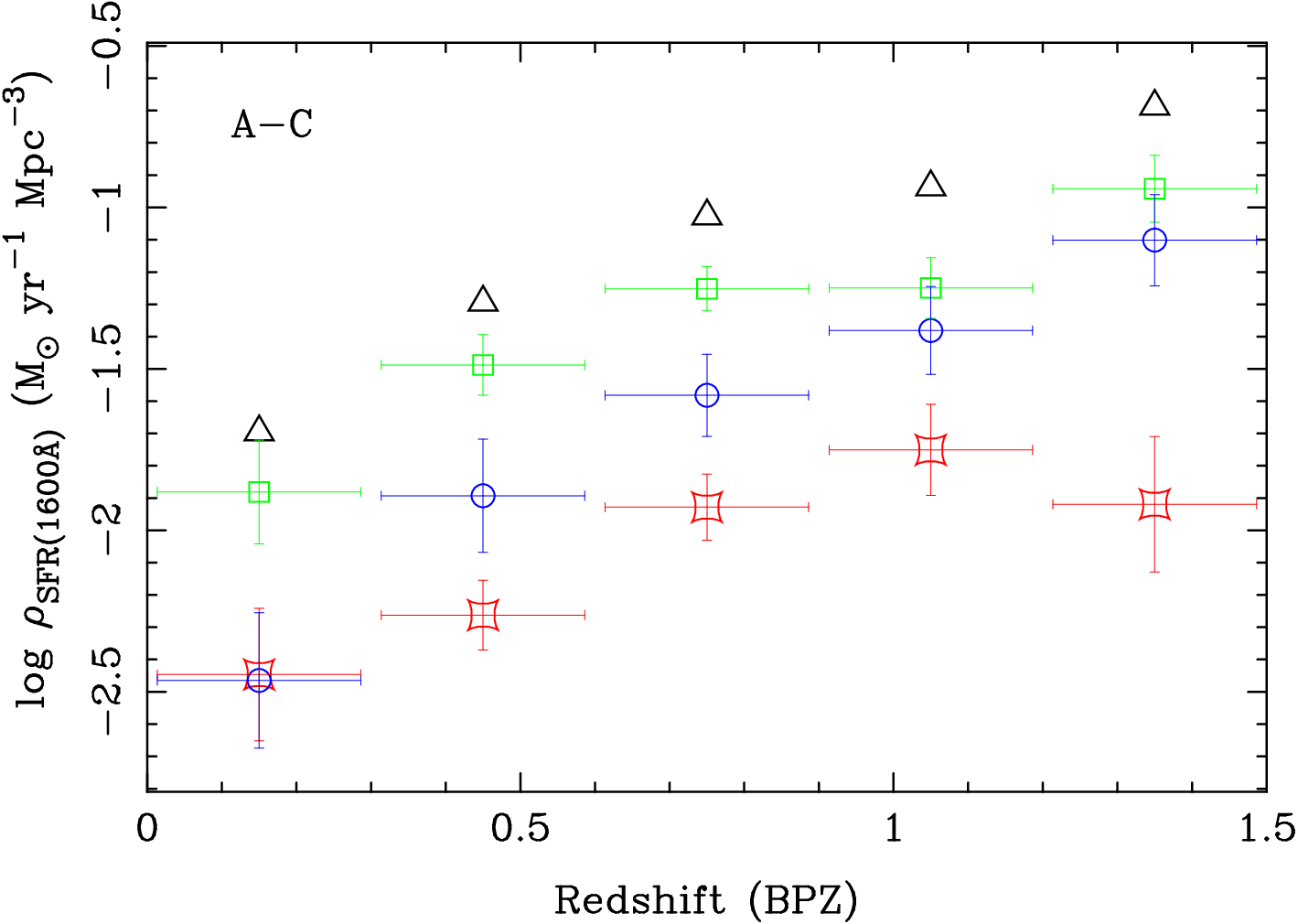}}
\caption{The co-moving star formation density of galaxies as a function
  of redshift for the three morphological groups. Red rounded squares
  represent E/S0s, green squares, Spirals and blue circles Irr/Pec
  selected galaxies. Bars represent $1\sigma$ deviations. Open
  triangles represent the total star formation density from all
  classes. The left panel shows the results from morphologically
  selecting galaxies according to their visual classes, while the
  right panel shows the results from the \AC\ automatic selection. }
\label{fig:sfr}
\end{figure*}

There is a plethora of studies that have focused on the evolution of
the SFR density, \rhoz, at intermediate redshifts
\citep{Gallego-etal-95,Madau-etal-96,Lilly-etal-96,
Flores-etal-99,Glazebrook-etal-04} and high redshifts
\citep{Giavalisco-etal-04b,Bouwens-etal-04, Coe-etal-05}. These studies
have found that SFR rises rapidly from $z=0$ to $z=1$ and peaks
somewhere at $z>2-3$ depending on the role of obscuration and our
ability to unearth missing galaxies at higher redshifts. 
However, there is little information about how the global SF history
is distributed into galaxy types (e.g. according to morphology). Only
a few attempts have been made to address this issue using relatively small,
but deep, sample such as the Hubble Deep Fields
\citep[see][]{Brinchmann-etal-98, Menanteau-etal-01}. In this paper we
aim to separate the contributions to the global SFR from each of the
morphological classes up to $z\simeq1$ . Many studies have addressed the
evolution of \rhoz\ in detail and at high redshifts, but a full
census of the contribution from individual morphological classes is
yet missing.

We calculate the SFR density for each morphological class at given
redshift interval $z_1<z<z_2$ as,
\begin{equation}
\rho_{\rm SFR}(z) = \sum_{j,z_1<z_j<z_2}\frac{{\rm SFR}_j}{V_{max,j}}
\end{equation}
where we sum the star formation contribution of each galaxy,
SFR$_{j}$, for $z_1<z_j<z_2$ and $V_{max}$ is the co-moving volume at
$z_{max}$, the highest redshift at which a given galaxy, $j$, is still
brighter than our sample limiting magnitude, \zbAB$<25$
\citep{Schmidt-68}. We proceed in this fashion for all galaxies, and
calculate \rhoz\ for the E/S0, Spirals and Irregular/Peculiar samples.
We explore the morphological selection calculating \rhoz\ using both
the visual morphological classifications and the machine-based
selection. In Fig.~\ref{fig:sfr} we show the results of \rhoz\ for
the visual and automatic classifications. We also show the total
\rhoz\ as well as the contribution of the three subclasses.  We note
that E/S0s are, as expected, the lowest contributors to the total
\rhoz\, regardless of the classification used.  We see from the figure
that there are only small differences in \rhoz\ from the two
approaches, mostly in the selection of E/S0 and Irr/Pec.
Moreover, the star formation density for early-types shows a modest
increase with redshift which seems to flatten out by $z\simeq1$. This
is not the case of either of the other classes or the global \rhoz. On the
other hand the Irr/Pec galaxies show a constant and in step increase
with redshift virtually matching the contribution from Spirals, the
most active group, by $z\simeq1$.

\subsection{Differential Evolution}

\begin{figure*}
\centerline{
\includegraphics[height=2.9in, angle=-90]{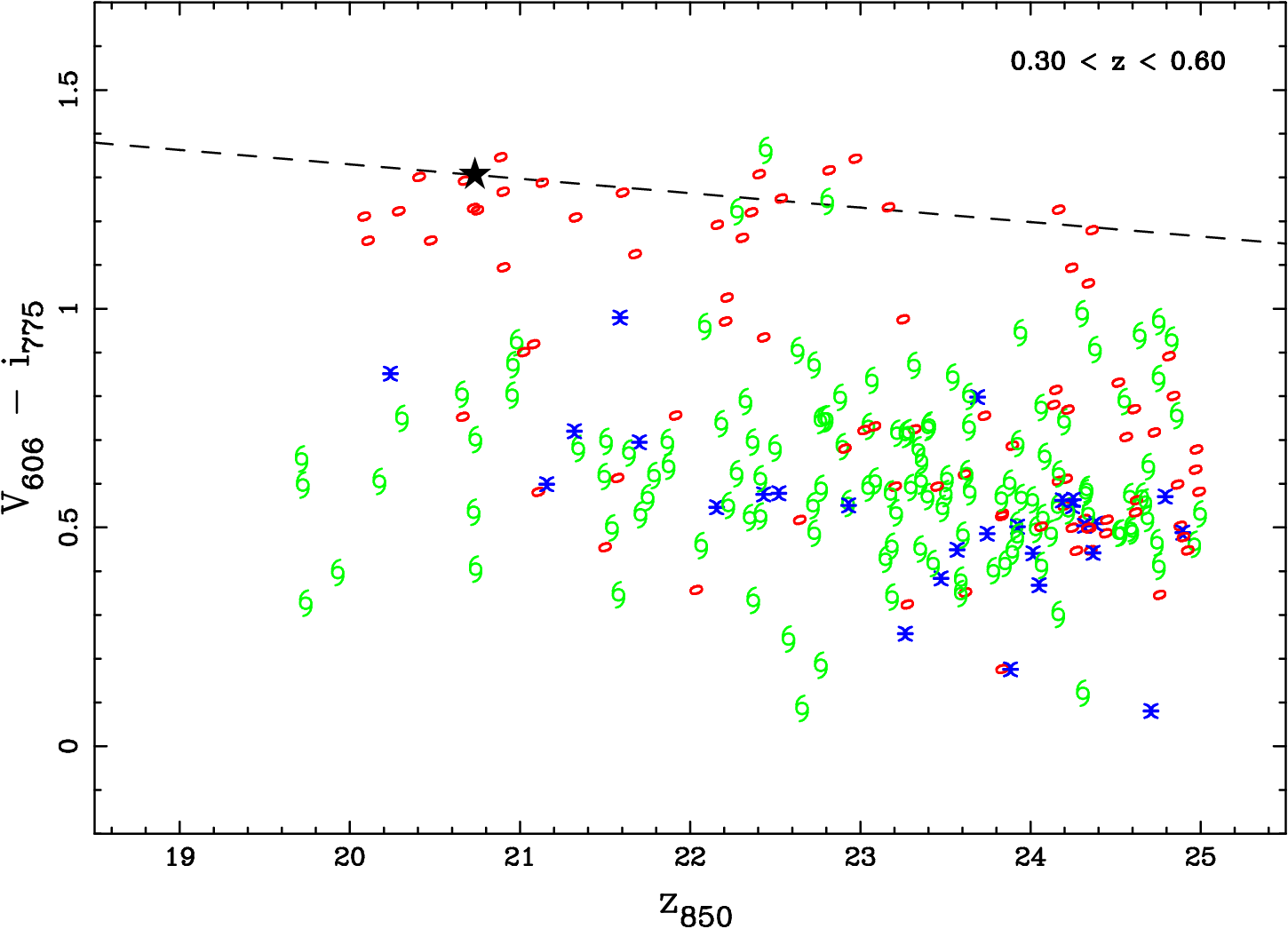}\hspace{0.5cm}
\includegraphics[height=2.9in, angle=-90]{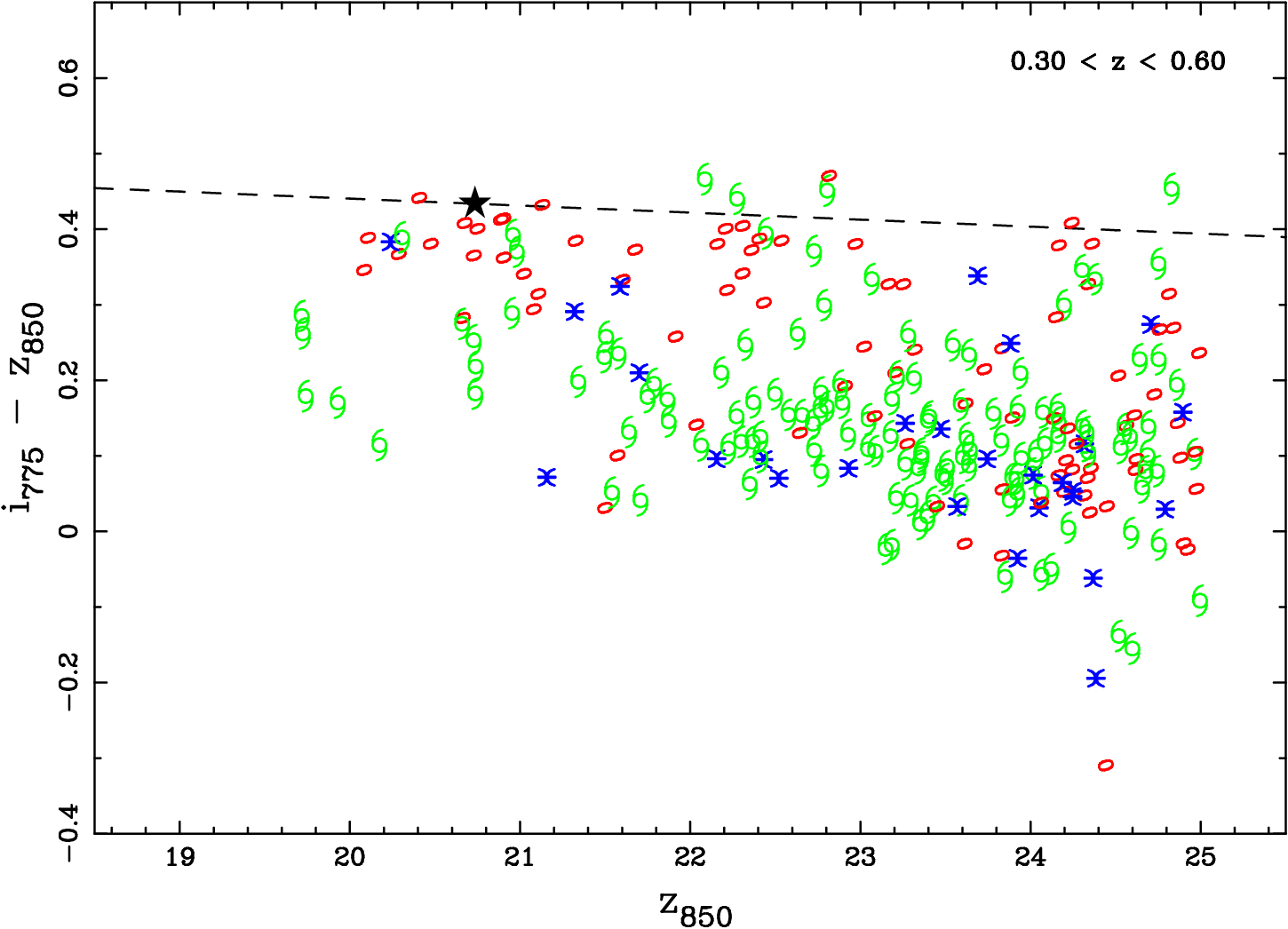}\\
}
\centerline{
\includegraphics[height=2.9in, angle=-90]{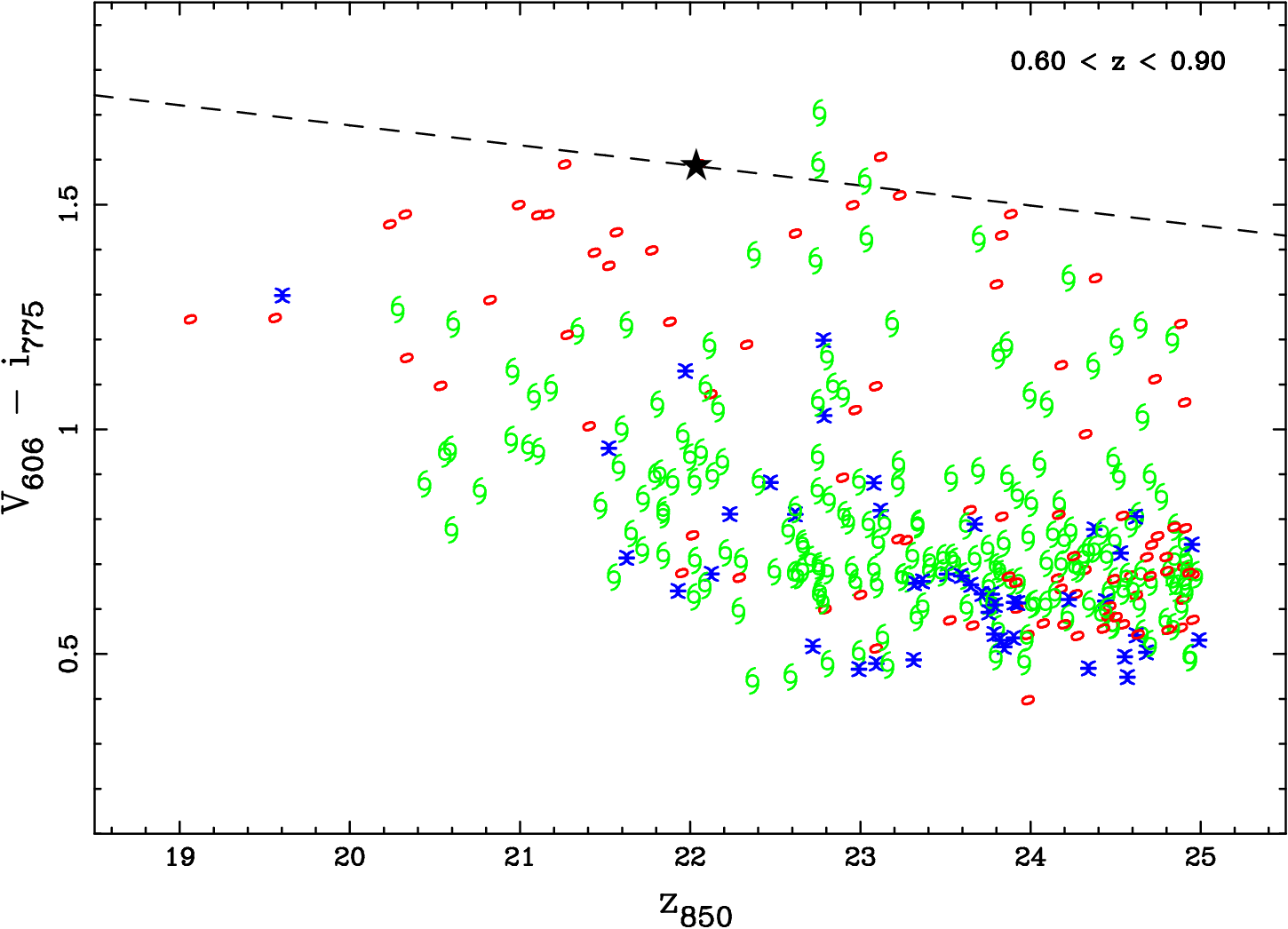}\hspace{0.5cm}
\includegraphics[height=2.9in, angle=-90]{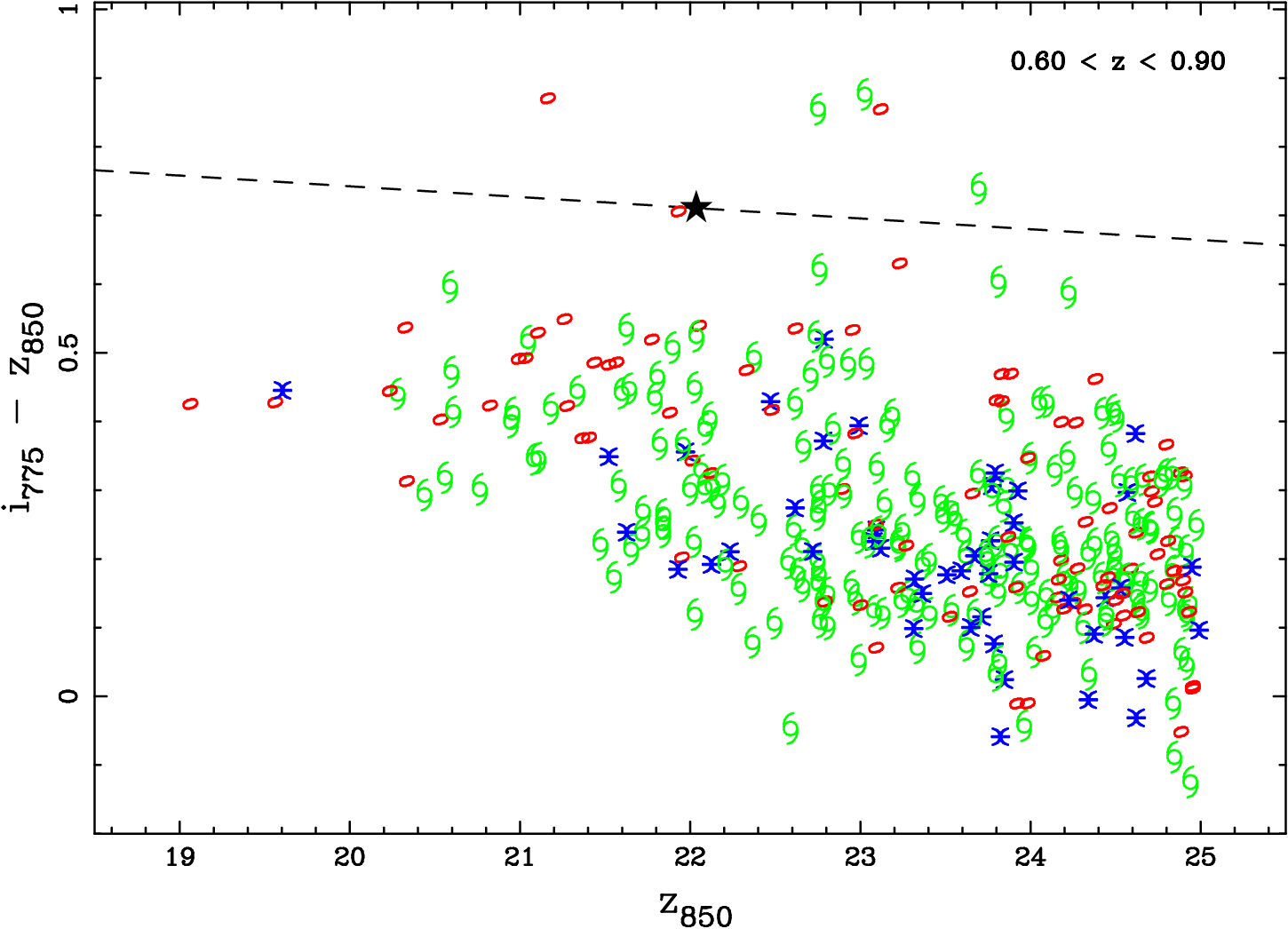}\\
}
\centerline{
\includegraphics[height=2.9in, angle=-90]{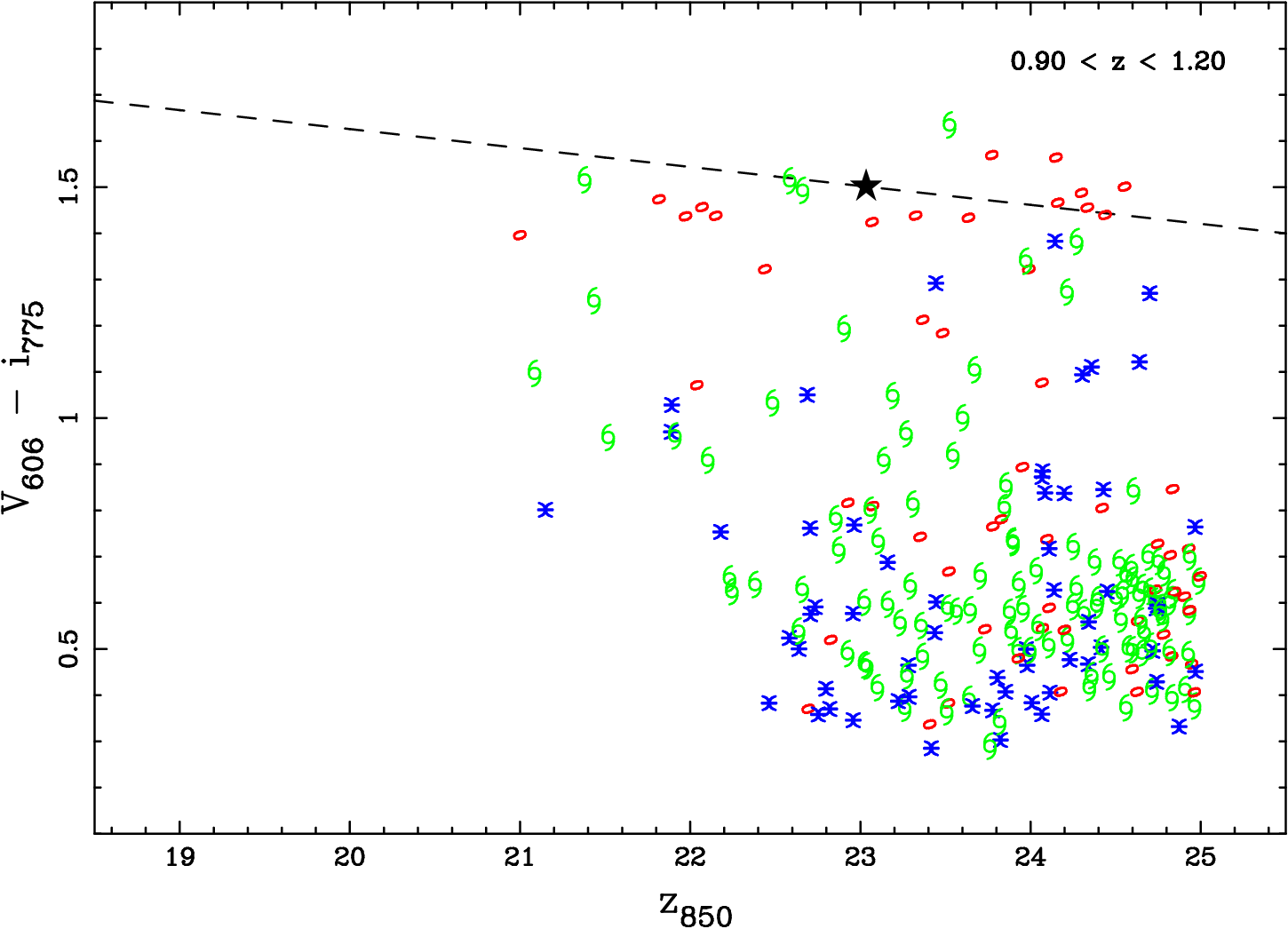}\hspace{0.5cm}
\includegraphics[height=2.9in, angle=-90]{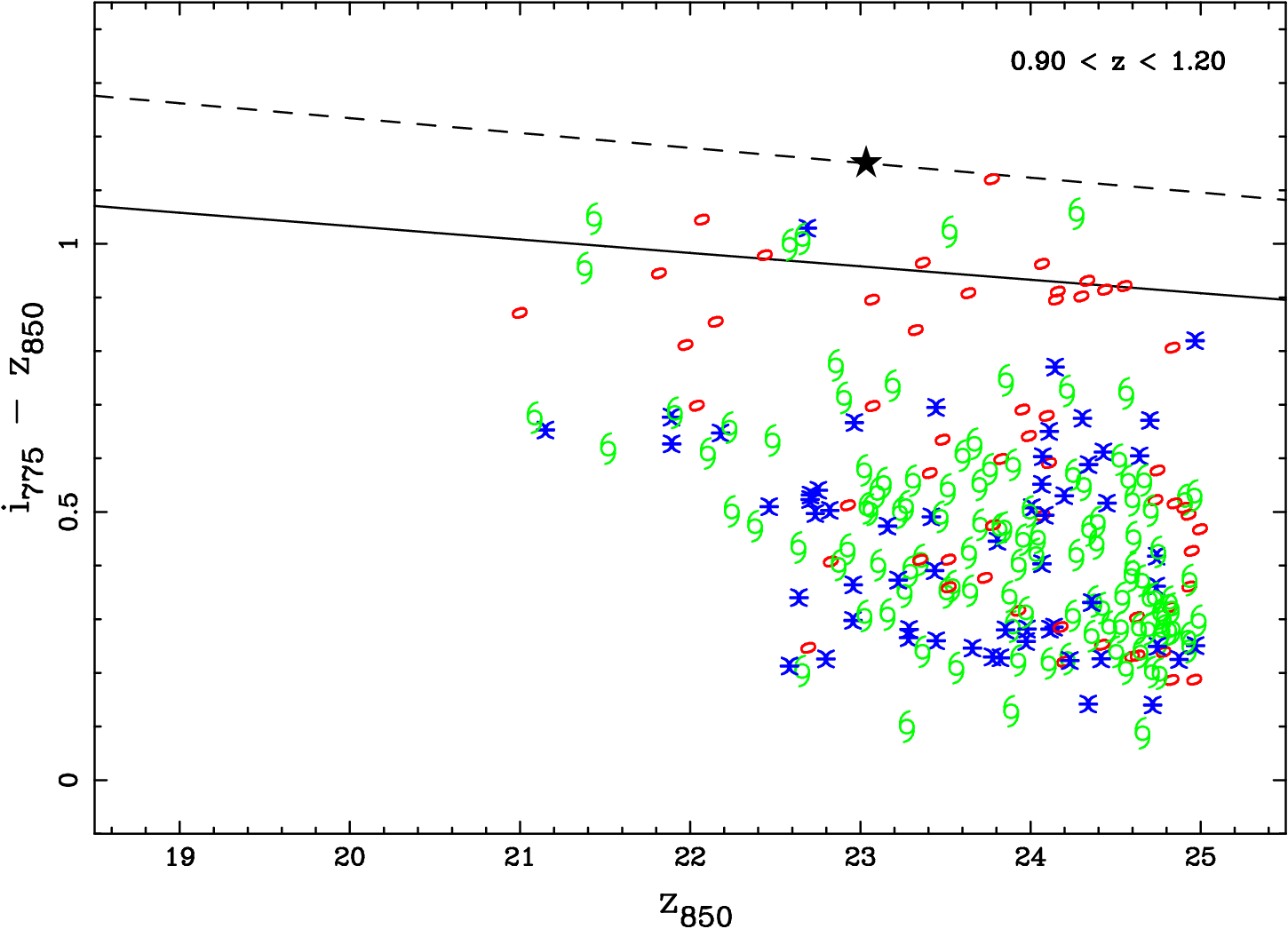}\\
}
\caption{The \Vi\ and \iz\ observed color magnitude diagrams as a
function of \zbAB\ magnitude for three evenly spaced redshift
intervals between $z=0.3-1.2$. Symbols are the same as in
Fig.~\ref{fig:eye-AC} and are keyed to their visual morphological
classes. E/S0 systems are represented by ellipses; Spiral galaxies by
spirals and Irr/Pec as asterisks. The dashed lines represent the color
magnitude relation for E/S0s for the Coma Cluster redshifted to mean
interval redshift of each panel. The solid line in the right panel is
the empirical relation for CL1252-2927 at $z=1.237$ from
\cite{Blakeslee-etal-03b}. For comparison we show as solid stars the
values of $M_{B}^{\star}$ from the COMBO17 survey \citep{COMBO17} for
early types redshifted to \zb\ at the mean redshift of each
interval.}
\label{fig:CMR}
\end{figure*}

We investigate the morphological evolution of galaxies using the
observed color magnitude diagrams at several redshift intervals. Our
motivation arises from the recognition that early-types in clusters
exhibit a tight correlation between their colors and magnitudes. This
correspondence, known as the color-magnitude relation (CMR) or ``red
sequence'', has been widely used to constrain the ages of galaxies,
particularly for cluster ellipticals \citep[see][and
references]{BLE92,Blakeslee-etal-03b}, and in recent years also for
field galaxies at different cosmic times
(\citealt{Bell-etal-04},\citealt*{KBB99}). Galaxies at similar
evolutionary stages should have comparable colors, while departures
from this can be employed to establish variations in their star
formation history and be interpreted as differential evolution within
their stellar population, regardless of its origin (i.e secular
evolution or recent mergers).

We construct the color magnitude (CM) diagrams for \iz\ and \Vi\
colors as a function of \zb\ at three evenly spaced redshift
intervals, $0.3<z<0.6$, $0.6<z<0.9$ and $0.9<z<1.2$. We chose to avoid
the $z<0.3$ objects because they are highly incomplete and
undersampled. Each interval's diagram is a snapshot of the galaxies'
current evolutionary phase, where dispersions in their color can be
interpreted as variations (or lack of) in their age and star formation
history. Despite the relative large width of our redshift bins, we can
use them to provide a general census of the galaxies' relative
formation stages as a function of their morphology.

In Fig~\ref{fig:CMR} we plot the diagrams keyed to their visual
morphological type, for \Vi\ in the left panel and \iz\ on the right
one. We use the same symbols as in Fig.~\ref{fig:eye-AC}. When
focusing on the red envelope of galaxies, we see that this is
dominated by morphologically selected E/S0 galaxies in all redshift
bins. Although there is considerable scatter, a red sequence of
passively evolved E/S0 is quite distinguishable. We also show the
$z=0.023$ Coma Cluster CMR \citep{BLE92} transformed to the intervals'
central redshift and transformed from $U-V$ to \iz$(z)$ and
\Vi$(z)$ accordingly. We use empirical and synthetic templates
\citep*{BC03,CWW-80} to transform between $\Delta(U-V)$ to
$\Delta(i_{775}-z_{850})$ and $\Delta(V_{606}-i_{775})$ respectively.
The redshifted Coma CMR represents a bona-fide benchmark for old and
coeval early-type systems. For the highest redshift \iz\ interval we
plot the empirical CMR for the $z=1.237$ CL1252-2927 cluster as
derived by \cite{Blakeslee-etal-03b}. In each of the panels we also
show with a solid star the redshifted values of $M_{B}^{\star}$ from
the COMBO17 survey \citep{COMBO17} to \zb\ at the central redshift of
each interval.
This insures that even in our highest redshift bin we are probing
galaxies at $\sim2$ magnitudes deeper than $M^{\star}$ within the flux
limit of our sample.

Previous CMR studies of HST morphologically selected field galaxies
show similar results. The \cite*{KBB99} analysis of ellipticals in the
Hubble Deep Field North at $\langle z \rangle\sim0.9$ was the first to
report the presence of a red sequence of field ellipticals at that
redshift, comprising around $\sim1/2$ of the population of
early types. However, a significant fraction have bluer rest-frame
colors, which \cite{Menanteau-etal-01} later reported to be also
coincidental with having large internal color dispersions. In a later
larger sample, \cite{Bell-etal-04} using GEMS observations confirmed
the presence of a red sequence dominated by early types ($70\%-80\%$)
at $z\sim0.7$. These studies establish a pattern in morphologically
selected samples of spheroids where systems are predominantly old and
coeval, but bluer and presumably younger ellipticals coexist at
intermediate redshifts \citep[see][]{Menanteau-etal-99,KBB99,
Abraham-etal-99, Menanteau-etal-01, Menanteau-etal-04, Benson-etal-02,
Cross-etal-04}.

In this paper we report a similar behavior. We note that while E/S0s
tend to dominate the red envelope, there is an important fraction of
systems with significantly bluer colors than those of old
ellipticals. Moreover, redder E/S0s tend to be brighter (massive) than
bluer ones, which are systematically fainter at all redshifts. This
result favors a view in which most massive ellipticals might have
formed at higher redshift and where at least an important fraction of
those were already in place at $z\simeq1$. However, it is unclear
whether bluer early-type galaxies will become the present day massive
systems we see in place. We note that it is unlikely that these bluer
systems could be confused with the brightest knots of faint irregular
galaxies, as it has been established that ACS can efficiently
distinguish between such contrasting galaxy types at these limits, and
especially in these high signal-to-noise images
\citep[see][]{Cross-etal-04,Conselice-etal-04,Menanteau-etal-04}.
Spiral galaxies display a broad range of
colors, and nearly uniform distribution of magnitudes overall redshift
ranges with observed magnitudes similar to those of the brightest
E/S0s in the red envelope. Irregular and Peculiar galaxies, on the
other hand, tend to dominate the low luminosity loci with
predominantly bluer colors, but comparable to the blue E/S0 systems. While
these two groups ---blue E/S0 and Irregular/Peculiars--- have
dramatically different structural properties, their similarities in
colors and luminosities make it tantalizing to relate the evolutionary
histories of these two very different types.

\section{Summary}

In this paper we have investigated the properties of a large
morphologically selected sample of ACS galaxies. The deep multicolor
information and ancillary ground-based spectroscopy has enabled us to
constrain the evolution of the SFR density and the individual
contribution to the global \rhoz. We report that spiral galaxies are
the main providers to the global SFR density at all redshifts, while
the E/S0s' humble contribution starts to flatten out by $z=1$. We note
that Irregular and Peculiar galaxies show the sharpest and continuous
rise with redshifts, reaching similar levels as spirals at
$z\simeq1.2$. We use the observed color magnitude diagram to constrain
the relative ages and formation histories of galaxies for different
morphologies. We report that E/S0s are the predominant population of
red and luminous galaxies at all redshift. We confirm, however, the
presence of a  population of much bluer and fainter early-type
galaxies, with colors and luminosities resembling those of the
irregular and peculiar galaxies. Upcoming deep far-ultraviolet
observations of the Ultra Deep Field will gives a clearer view of the
star-forming nature of galaxies in the UV window.

\acknowledgments

ACS was developed under NASA contract NAS 5-32865, and this research
is supported by NASA grant NAG5-7697. We would like to thank Alex
Framarini and Ken Anderson for their valuable technical support to the
ACS Science Team.


\end{document}